\documentclass[11pt,a4paper]{article}

\usepackage[utf8]{inputenc}
\usepackage[T1]{fontenc}
\usepackage{amsmath,amssymb,amsfonts}
\usepackage{graphicx}
\usepackage{multirow}
\usepackage{subcaption}
\usepackage{algorithm2e}
\usepackage[hidelinks,colorlinks=true,linkcolor=blue,citecolor=blue,urlcolor=blue]{hyperref}
\usepackage{booktabs}
\usepackage{xcolor}
\usepackage{url}
\usepackage[margin=1in]{geometry}
\usepackage{natbib}

\usepackage{float}
\title{\textbf{RefineFormer3D: Efficient 3D Medical Image Segmentation via Adaptive Multi-Scale Transformer with Cross Attention Fusion}}

\author{
\textbf{Kavyansh Tyagi} \\
National Institute of Technology Kurukshetra \\
Department of Electronics and Communication Engineering \\
\texttt{123109026@nitkkr.ac.in}
\and
\textbf{Vishwas Rathi$^*$} \\
National Institute of Technology Kurukshetra \\
Department of Computer Engineering \\
\texttt{vishwas@nitkkr.ac.in}
\and
\textbf{Puneet Goyal} \\
Indian Institute of Technology Ropar \\
Department of Computer Science and Engineering \\
\texttt{puneet@iitrpr.ac.in}
}

\date{}

\begin{document}

\maketitle

\vspace{-0.3cm}
\begin{center}
\small $^*$ Corresponding author
\end{center}

\begin{abstract}
Accurate and computationally efficient 3D medical image segmentation remains a critical challenge in clinical workflows. Transformer-based architectures often demonstrate superior global contextual modeling but at the expense of excessive parameter counts and memory demands, restricting their clinical deployment. We propose RefineFormer3D, a lightweight hierarchical transformer architecture that balances segmentation accuracy and computational efficiency for volumetric medical imaging. The architecture integrates three key components: (i) GhostConv3D-based patch embedding for efficient feature extraction with minimal redundancy, (ii) MixFFN3D module with low-rank projections and depthwise convolutions for parameter-efficient feature extraction, and (iii) a cross-attention fusion decoder enabling adaptive multi-scale skip connection integration. RefineFormer3D contains only 2.94M parameters, substantially fewer than contemporary transformer-based methods. Extensive experiments on ACDC and BraTS benchmarks demonstrate that RefineFormer3D achieves 93.44\% and 85.9\% average Dice scores respectively, outperforming or matching state-of-the-art methods while requiring significantly fewer parameters. Furthermore, the model achieves fast inference (8.35 ms per volume on GPU) with low memory requirements, supporting deployment in resource-constrained clinical environments. These results establish RefineFormer3D as an effective and scalable solution for practical 3D medical image segmentation.

\end{abstract}

\section{Introduction}
\label{sec:intro}

Deep learning has fundamentally reshaped medical image analysis as extraction of complex patterns from clinical data can be done automatically at unprecedented scales. Among these advances, 3D medical image segmentation stands as a foundational task which is crucial for applications ranging from organ localization and tumor delineation to treatment planning. Traditional encoder-decoder networks such as U-Net~\cite{ronneberger2015u} and its derivatives~\cite{isensee2021nnu,zhou2018unet++} have long been the backbone of volumetric segmentation, and they effectively compress input data into latent representations and reconstruct dense voxel-wise predictions. However, the limited receptive field of convolutional operators and their inherent locality bias restrict their capacity to model global anatomical context, particularly in cases involving large inter patient variation in scale, texture, and shape.

To address these limitations, transformer based architectures have emerged as a powerful alternative as they leverage global self-attention mechanisms~\citep{hussain2022global} to capture long range dependencies and semantic coherence across medical volumes. Pioneering contributions such as TransUNet~\citep{chen2021transunettransformersmakestrong} and UNETR~\citep{UNETR} have demonstrated the effectiveness of integrating Vision Transformers with convolutional decoders, leading to substantial improvements over earlier purely convolutional approaches. SWIN-Unet~\citep{swinUNet} have further explored hierarchical and pure transformer U-shaped architectures, affirming the value of transformer backbones for contextual representation learning in segmentation tasks.

However, these gains come at a cost. As full self-attention incurs heavy memory overhead and computational burden, it raises concerns about clinical feasibility where resource constraints matter. This limits their applicability in real world clinical scenarios where efficiency and reliability are paramount. Moreover, the current prevailing skip fusion strategies are typically based on static concatenation or convolutional operations, and they may inadequately integrate multi-scale features, which undermines segmentation performance in anatomically complex or ambiguous regions. The conventional concatenation-based skip connections treat all encoder features uniformly, failing to selectively aggregate semantically relevant information for the decoder's current reconstruction state. This naive fusion not only introduces redundant features but also foregoes adaptive, query driven selection mechanisms that could align encoder context with decoder specific requirements. This limitation becomes acute when segmenting the heterogeneous anatomical structures with variable appearances.

Recent state-of-the-art research, including nnFormer~\citep{nnFormer} and SegFormer3D~\citep{10678245}, tried to reconcile this trade-off between performance and efficiency by introducing hybrid or windowed attention schemes and lighter multilayer perceptron (MLP) based decoders. These architectures have made strides in reducing computational costs. Models such as LeViT-UNet~\citep{LeViT-UNet} have specifically targeted inference efficiency by employing fast transformer encoders. Despite these advances, many contemporary transformer models still retain excessive parameter counts, especially within skip fusion and decoding modules, or they tend to sacrifice global feature integration to achieve efficiency. Furthermore, the repetitive and compressible nature of volumetric medical data a characteristic ideally suited for lightweight, context aware modeling remains underexploited in existing segmentation frameworks.

\begin{figure}[!t]
\centering
\includegraphics[width=0.85\columnwidth]{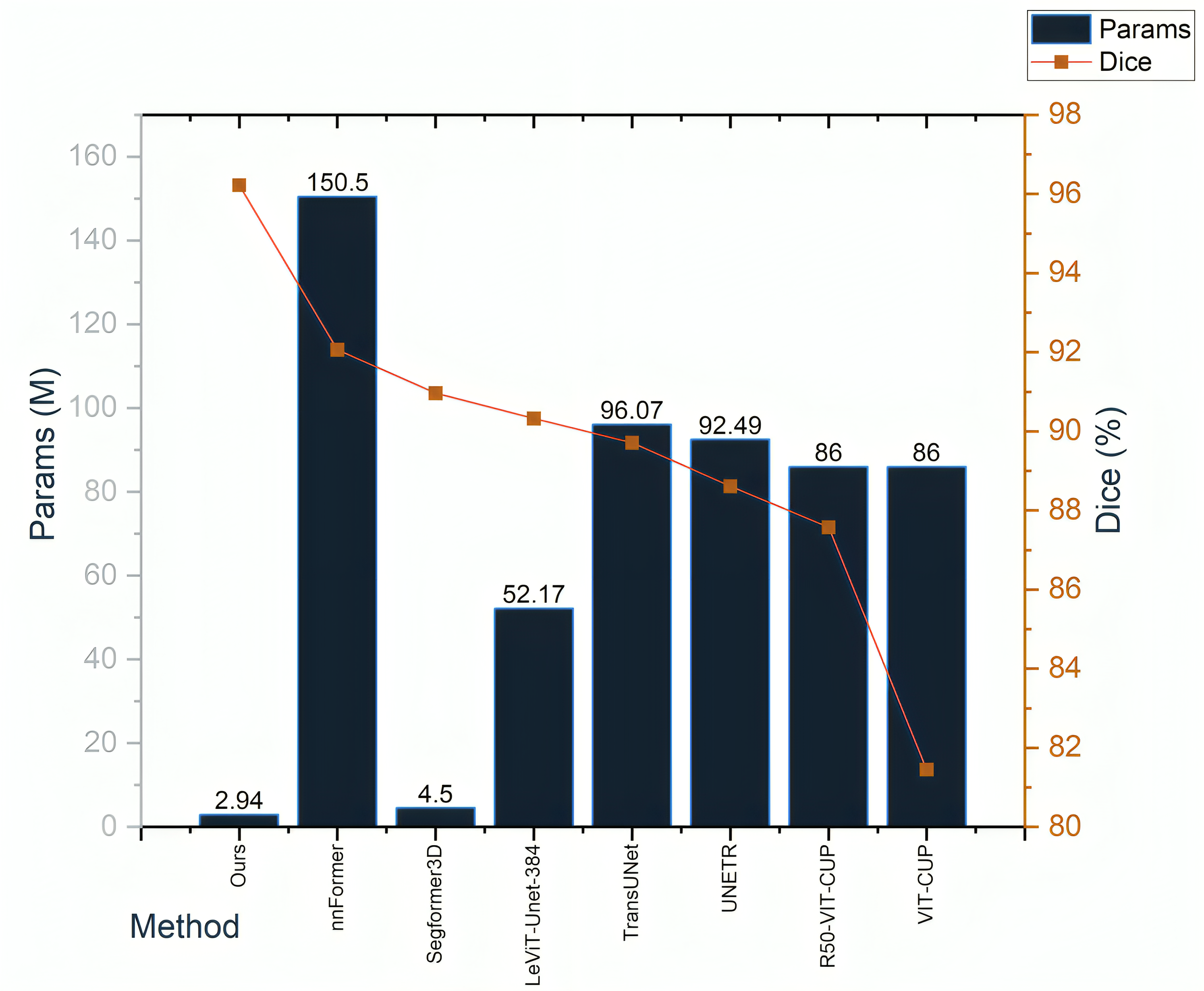}
\caption{Model efficiency comparison on ACDC dataset~\citep{acdc}. Parameter count versus segmentation performance for RefineFormer3D and existing 3D segmentation models. Blue bars indicate number of parameters; orange-red curve shows average Dice score. RefineFormer3D achieves superior performance with only 2.94M parameters.}
\label{fig:efficiency_chart}
\end{figure}

In response to these persistent challenges, we propose RefineFormer3D, a hierarchical multi-scale transformer architecture engineered for both parametric efficiency and robust contextual reasoning in 3D medical image segmentation. The core of our method is the cross attention fusion decoder block, which uses GhostConv3D for efficient 3D feature processing and enhances it with channel wise attention using Squeeze and Excitation mechanisms. This adaptively and dynamically aggregates the multi-scale features across the decoder. The attention aware fusion strategy refines semantic integration throughout the network while minimizing computational overhead. The encoder employs hierarchical windowed self-attention and \\ MixFFN3D modules with low-rank projections and depthwise 3D convolutions. This design captures both global dependencies and fine anatomical details, effectively balancing accuracy and computational efficiency. 

Figure~\ref{fig:efficiency_chart} illustrates the performance efficiency trade-off on the ACDC dataset~\citep{acdc}. RefineFormer3D achieves superior performance with only 2.94 million parameters, outperforming state-of-the-art methods and demonstrating its effectiveness as a 
lightweight yet powerful 3D segmentation architecture.

Extensive experiments on widely used medical segmentation benchmarks, including ACDC~\citep{acdc}, and BraTS~\citep{4b589b6824a64a2a91e8e3b26cc0bf9e}, demonstrate that RefineFormer3D not only outperforms state-of-the-art models such as nnFormer, SegFormer3D, and UNETR in segmentation accuracy but also achieves notable reductions in parameter count and inference time. These results project RefineFormer3D as a strong candidate for clinical deployment, which meets the critical requirements of accuracy, reliability and computational efficiency. The main contributions of this work are: (i) a hierarchical transformer architecture for efficient 3D medical image segmentation with only 2.94M parameters, (ii) an adaptive decoder block that dynamically integrates multi-scale features using attention guided fusion mechanisms, (iii) a comprehensive evaluation on two benchmark datasets demonstrating superior performance efficiency trade-offs compared to existing methods, and (iv) detailed ablation studies validating the contribution of each architectural component.

\section{Related Work}
\label{sec:rel}

Recent advances in 3D medical image segmentation explore CNNs, transformers and their hybrids to improve performance and efficiency. While CNNs excel at capturing local features, transformers enhance global context understanding through self-attention mechanisms. We divide the existing works into four categories: (i) CNN-based approaches, (ii) Pure transformer based methods, (iii) Hybrid CNN-Transformer architectures and (iv) Efficient transformer variants.

\subsection{CNN-Based Approaches}

The U-Net architecture~\citep{ronneberger2015u} and its many descendants, including deep-supervised CNNs~\citep{33/7965852}, DenseUNet~\citep{2/QIMS43519}, and 3D variants such as 3D U-Net~\citep{6/çiçek20163dunetlearningdense}, V-Net~\citep{21/7785132}, and nnUNet~\citep{13/isensee2018nnunetselfadaptingframeworkunetbased}, have driven significant progress in medical image segmentation. Convolutional networks provide effective multi-scale feature extraction but their fundamentally local receptive fields limit their ability to accurately segment complex and context-dependent structures, particularly in 3D volumes.

CNN-based improvements, such as dilated convolutions, attention gates, and pyramid pooling, have enhanced context modeling~\citep{chen2024frequency, wu2022p2t, chen2021end}. However, these approaches remain constrained by the inherent locality of convolution operations, often failing to model long-range dependencies or capture subtle anatomical variations. Attempts to mitigate these weaknesses through deeper or wider architectures~\citep{2a, 3a} frequently result in increased computational demands without proportional gains in accuracy.

\subsection{Pure Transformer Architectures}

SETR~\citep{zheng2021rethinkingSETR} and other early transformer-based architectures reformulated segmentation as a sequence-to-sequence prediction task, employing pure transformers as encoders for global context. However, these models often suffer from poor localization due to the loss of spatial detail during patch embedding and the absence of hierarchical, multi-resolution features, requiring complex or inefficient decoders for spatial recovery.

Transformer-only architectures such as Swin-UNet~\citep{swinUNet} and SwinUNETR~\citep{10.1007/978-3-031-08999-2_22} leveraged hierarchical attention with shifted windows to efficiently learn local-to-global representations. However, their dependence on windowed attention may restrict context modeling for large or irregular anatomical structures and contributes to relatively high parameter counts. Recent lightweight designs such as GCI-Net~\citep{10778607} and PFormer~\citep{GAO2025107154} addressed this by introducing efficient global context modules and content-driven attention, maintaining strong performance at a fraction of traditional transformer complexity.

\subsection{Hybrid CNN-Transformer Architectures}

Hybrid CNN-transformer architectures, such as TransBTS~\citep{TransBTS}, CoTr~\citep{DBLP:journals/corr/abs-2012-15840}, TransFuse~\citep{zhang2021transfuse}, and LoGoNet~\citep{logonetkarimi2024masked}, have been proposed to leverage the complementary strengths of both paradigms. These models incorporated diverse fusion and attention strategies, including deformable attention~\citep{29/xia2022vision} and explicit integration of CNN and transformer features~\citep{zhang2021transfuse}, to enhance segmentation performance. While these hybrid models effectively capture both local and global context, they often involve complex architectures that can lead to slower inference and limited scalability when applied to larger or more diverse clinical datasets.

Recent works such as DAUNet~\citep{LIU202558}, DS-UNETR++~\citep{Jiang2025DSUNETRPP}, and MS-TCNet~\citep{AO2024108057} have further refined this paradigm by introducing deformable aggregation, gated dual-scale attention, and multi-scale transformer-CNN fusion, achieving higher Dice accuracy with competitive parameter counts.

\subsection{Efficient Transformer Variants}

Efficient transformer variants such as LeViT-UNet~\citep{LeViT-UNet} and SegFormer3D~\citep{10678245} further addressed the high computational and parameter complexity by integrating lightweight transformer modules and streamlined decoders, such as all-MLP designs. LeViT-UNet introduced a fast hybrid encoder but was limited by shallow multi-scale fusion, whereas SegFormer3D demonstrated strong efficiency and competitive accuracy by employing hierarchical attention and a lightweight decoder with 4.51M parameters. Despite these gains, SegFormer3D and its contemporaries continue to rely on largely static or parameter-heavy skip fusion strategies, which can fail to adaptively exploit anatomical variation or handle challenging low-contrast targets.

Methods including nnFormer~\citep{nnFormer}, which employed skip attention instead of concatenation for feature fusion, and UNETR~\citep{UNETR}, which used a transformer as the primary encoder for 3D segmentation, demonstrated improved segmentation performance and better global context modeling. However, these methods often remain computationally demanding and do not fully resolve the tension between fine-scale localization and adaptive, efficient feature integration.

\begin{table}[!t]
\centering
\caption{Comparison of RefineFormer3D with state-of-the-art methods in terms of model size and computational complexity. RefineFormer3D demonstrates notable reduction in parameters while maintaining competitive computational cost.}
\label{table1}
\small
\begin{tabular}{@{}lrr@{}}
\toprule
\textbf{Architecture} & \textbf{Params (M)} \\
\midrule
nnFormer~\citep{nnFormer} & 150.5 \\
TransUNet~\citep{chen2021transunettransformersmakestrong} & 96.07  \\
UNETR~\citep{UNETR} & 92.49  \\
DS-UNETR++~\citep{Jiang2025DSUNETRPP} & 67.7  \\
SwinUNETR~\citep{10.1007/978-3-031-08999-2_22} & 62.83  \\
MS-TCNet~\citep{AO2024108057} & 59.49  \\
PFormer~\citep{GAO2025107154} & 46.04  \\
DAUNet~\citep{LIU202558} & 16.36  \\
GCI-Net~\citep{10778607} & 13.36  \\
SegFormer3D~\citep{10678245} & 4.51  \\
\midrule
\textbf{RefineFormer3D} & \textbf{2.94}  \\
\bottomrule
\end{tabular}
\end{table}

Table~\ref{table1} presents a comprehensive comparison of RefineFormer3D with state-of-the-art methods in terms of parameter count. Despite advances in recent architectures, most current models employ rigid or parameter-heavy skip fusion schemes that limit their adaptability to anatomical differences and hinder both segmentation accuracy and efficiency. RefineFormer3D addresses these limitations by introducing adaptive, attention-based skip fusion within a lightweight transformer design, achieving superior accuracy and resource efficiency compared to previous methods.

\section{Proposed Methodology}
\label{sec:method}

\begin{figure*}[!t]
\centering
\includegraphics[width=0.95\textwidth]{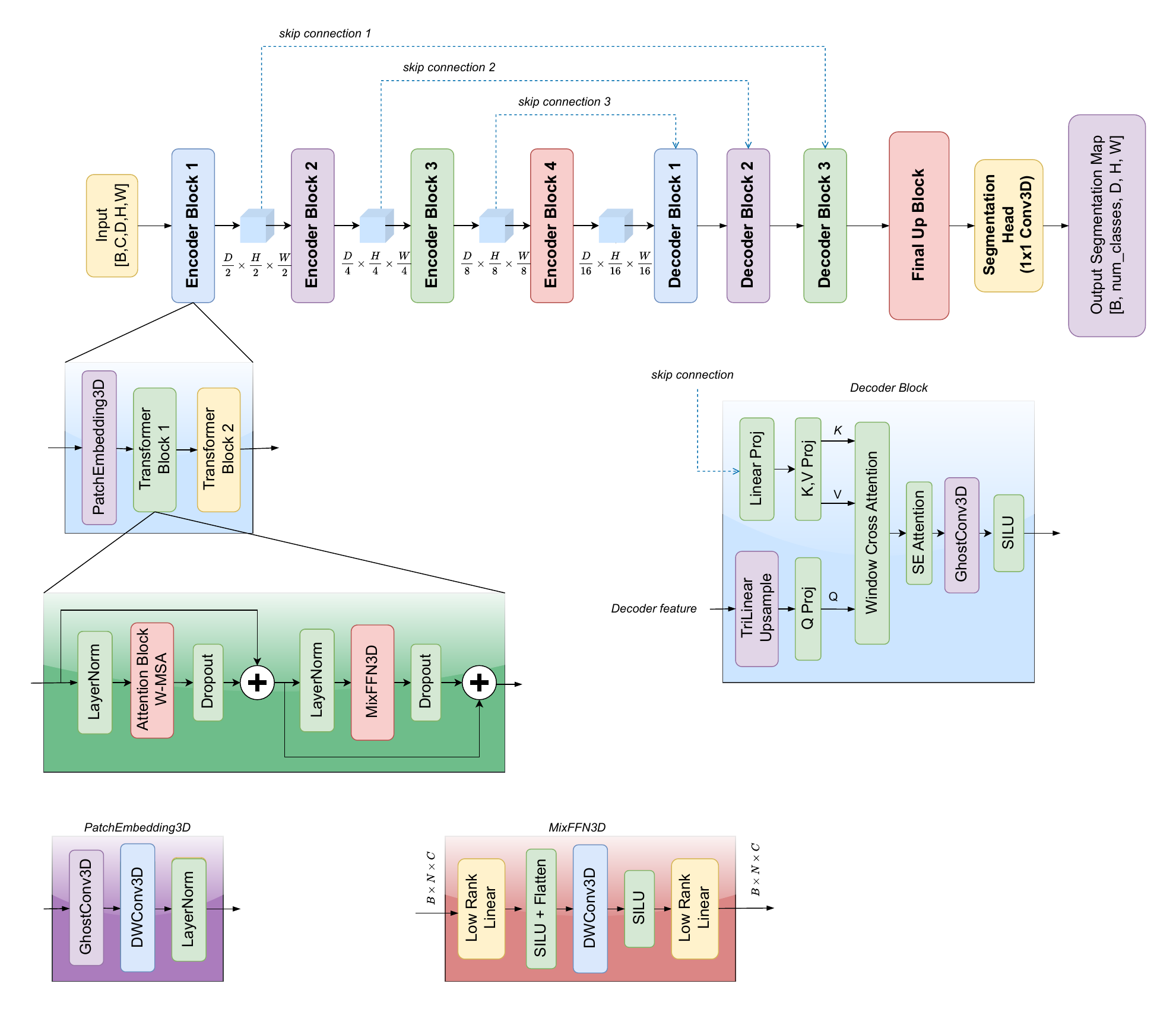}

\caption{RefineFormer3D architecture overview. The input volume \(\mathbb{R}^{B \times C_{\mathrm{in}} \times D \times H \times W}\) is procecssed via 3D patch embedding and encoded through four hierarchical stages. Each encoder stage applies two consecutive transformer blocks which are windowed self attention followed by shifted window attention. The decoder progressively refines features by fusing encoder skip connections through cross attention, where decoder features query encoder representations for selective multi scale aggregation. The final upsampling block performs refinement without skip connections. Deep supervision is applied via auxiliary heads at intermediate decoder outputs.}

\label{fig:refineformer3d}
\end{figure*}

RefineFormer3D is designed to deliver state-of-the-art segmentation accuracy for volumetric medical images while maintaining exceptional computational efficiency. The architecture requires only 2.94 million parameters, an order of magnitude less than leading transformer based baselines. This section outlines the key architectural components introduced in the encoder and decoder modules. The proposed RefineFormer3D architecture is illustrated in Figure~\ref{fig:refineformer3d}.

\subsection{Encoder Architecture}

The encoder in RefineFormer3D efficiently extracts hierarchical features from 3D medical images through a multi-stage processing pipeline. It addresses memory and computational limitations of previous transformer-based segmentation models, enabling better scalability for clinical deployment. The encoder comprises three key components working in sequence: (i) GhostConv3D~\citep{han2020ghostnet} based patch embedding for efficient low-level feature extraction, (ii) Transformer blocks that progressively capture multi-scale context through shifted window attention, (iii) MixFFN3D module that perform parameter-efficient feature mixing using low-rank projections and depthwise convolutions. It also uses hybrid normalization with stochastic depth regularization to ensure stable training. Each component of encoder is designed to balance computational efficiency with representational power, as detailed below.

\subsubsection{PatchEmbedding3D}

Input volumes are first processed by a GhostConv3D based patch embedding layer that preserves local voxel continuity while using substantially fewer parameters and computations than conventional 3D convolutions. Unlike standard patch embedding which often ignores spatial redundancy, GhostConv3D generates a set of primary feature maps via regular convolution and then augments them with ``ghost'' features using lightweight depthwise convolutions. To the best of our knowledge, this represents the first application of GhostConv3D for 3D transformer patch embedding in medical imaging.

Given input volume \(\mathbf{X} \in \mathbb{R}^{B \times C_{\mathrm{in}} \times D \times H \times W}\), where \(B\) represents the batch size, \(C_{\mathrm{in}}\) is the number of input channels, and \(D\), \(H\), \(W\) represent depth, height, and width respectively, the GhostConv3D patch embedding computes:
\begin{align}
\mathbf{F}_{\mathrm{primary}} &= \mathrm{Conv3D}(\mathbf{X}; \mathbf{W}_{\mathrm{prim}}) \\
\mathbf{F}_{\mathrm{ghost}} &= \mathrm{DWConv3D}(\mathbf{F}_{\mathrm{primary}}; \mathbf{W}_{\mathrm{ghost}}) \\
\mathbf{F}_{\mathrm{out}} &= \mathrm{Concat}\left[ \mathbf{F}_{\mathrm{primary}}, \mathbf{F}_{\mathrm{ghost}} \right]
\end{align}

where \(\mathbf{W}_{\mathrm{prim}}\) and \(\mathbf{W}_{\mathrm{ghost}}\) represent the learnable weights of the primary and depthwise Conv3D layers, respectively. Here, \(\mathbf{F}_{\mathrm{primary}} \in \mathbb{R}^{B \times C_1 \times D' \times H' \times W'}\), \(\mathbf{F}_{\mathrm{ghost}} \in \\ \mathbb{R}^{B \times (C_{\mathrm{out}} - C_1) \times D' \times H' \times W'}\) and \(\mathbf{F}_{\mathrm{out}} \in \mathbb{R}^{B \times C_{\mathrm{out}} \times D' \times H' \times W'}\). The parameter \(C_1 = \lfloor C_{\mathrm{out}} / r \rfloor\) where \(r\) is the ghost ratio (typically \(r = 2\)), and \(C_\mathrm{out}\) is the output channel size. The depthwise convolution \(\mathrm{DWConv3D}\) operates with kernel size \(3^3\) and applies a separate convolutional filter for each input channel, producing ghost features that capture local spatial variations with minimal additional parameters (\(27C_1\) compared to \(C_1 \cdot C_{\mathrm{out}} \cdot 27\) for standard convolution). The output spatial dimensions \(D'\), \(H'\), \(W'\) are computed as \(S' = \lfloor \frac{S + 2p - k}{s} \rfloor + 1\) for each axis, where \(S\) is the input size, \(k\) the kernel size, \(s\) the stride, and \(p\) the padding.

The parameter reduction achieved by GhostConv3D can be quantified as follows: while standard Conv3D requires \(C_{\mathrm{in}} \cdot C_{\mathrm{out}} \cdot k^3\) parameters, GhostConv3D requires only \(C_{\mathrm{in}} \cdot C_1 \cdot k^3 + (C_{\mathrm{out}} - C_1) \cdot k^3 \approx C_{\mathrm{in}} \cdot C_{\mathrm{out}} \cdot k^3 / r\), yielding approximately 2\(\times\) parameter reduction for \(r=2\).

Before flattening to tokens, the output passes through a 3D positional convolution (PosConv) block followed by LayerNorm, which together embed spatial location information and normalize feature distributions. The PosConv block employs a depthwise 3D convolution with kernel size \(3^3\) and zero padding to encode relative spatial positions:
\begin{equation}
\mathbf{F}_{\mathrm{pos}} = \mathrm{DWConv3D}(\mathbf{F}_{\mathrm{out}}; \mathbf{W}_{\mathrm{pos}}, k{=}3, \text{groups}{=}C_{\mathrm{out}})
\end{equation}
where each channel learns its own positional encoding pattern. This is followed by LayerNorm to stabilize training:
\begin{equation}
\mathbf{F}_{\mathrm{norm}} = \mathrm{LayerNorm}(\mathbf{F}_{\mathrm{pos}})
\end{equation}
The normalized features are then flattened along spatial dimensions to produce token sequence \(\mathbf{T} \in \mathbb{R}^{B \times N \times C_{\mathrm{out}}}\) where \(N = D' \times H' \times W'\). These spatially aware tokens then serve as input to the hierarchical transformer stages for multi-scale feature extraction.

\subsubsection{Transformer Block }

The PatchEmbedding3D output is processed through two sequential transformer blocks having distinct attention mechanisms. The first block computes self attention within fixed spatial windows capturing local context efficiently. The second block employs shifted window partitioning to establish inter window connections which addresses the limitation of isolated local receptive fields. The non overlapping windowed self-attention mechanisms~\citep{feng2024window} capture long range dependencies within local volumetric regions balancing global context modeling with computational efficiency. Following the Swin Transformer~\citep{liu2021swin} alternate transformer blocks use shifted window partitioning (shift by $\lfloor w/2 \rfloor$) to enable cross-window information exchange. Each of the encoder stage applies two consecutive transformer blocks- the first with regular window partitioning and the second with shifted windows which creates a pattern that facilitates information flow across window boundaries.

Given input features $\mathbf{X} \in \mathbb{R}^{B \times D \times H \times W \times C}$, where $D$, $H$, $W$ denote the spatial dimensions and $C$ the channel dimension, a cyclic shift by $\vec{s} = (s_d, s_h, s_w)$ voxels (where $s_d, s_h, s_w$ are shift amounts along depth, height, and width, typically set to $(\lfloor w_d/2 \rfloor, \lfloor w_h/2 \rfloor, \lfloor w_w/2 \rfloor)$ for shifted blocks and $(0,0,0)$ for regular block) is applied in alternate blocks, followed by partitioning into $M$ windows of size $(w_d, w_h, w_w)$:

\begin{align}
\mathbf{X}' &= \mathrm{CyclicShift}(\mathbf{X}, \vec{s}) \\
\{\mathbf{X}^{(m)}\} &= \mathrm{Partition}(\mathbf{X}', (w_d, w_h, w_w)), \quad m=1,\ldots, M
\end{align}
where $M = \frac{D \cdot H \cdot W}{w_d \cdot w_h \cdot w_w}$ represents the total number of windows. Each window $\mathbf{X}^{(m)} \in \mathbb{R}^{B \times (w_d \cdot w_h \cdot w_w) \times C}$ is treated as a sequence of $w_d \cdot w_h \cdot w_w$ tokens for self-attention computation.

Within each window, we compute standard multi-head self-attention with relative positional bias:
\[
\mathrm{Attention}(\mathbf{Q}, \mathbf{K}, \mathbf{V}) = \mathrm{Softmax}\left(\frac{\mathbf{Q}\mathbf{K}^\top}{\sqrt{d_k}} + \mathbf{B}\right)\mathbf{V}
\]
where \(\mathbf{Q} = \mathbf{X}^{(m)}\mathbf{W}_Q\), \(\mathbf{K} = \mathbf{X}^{(m)}\mathbf{W}_K\), \(\mathbf{V} = \mathbf{X}^{(m)}\mathbf{W}_V\) are projections of the window tokens with learnable weight matrices $\mathbf{W}_Q, \mathbf{W}_K, \mathbf{W}_V \in \mathbb{R}^{C \times d_k}$, and $d_k = C/H_{\text{num}}$ is the dimension per attention head with $H_{\text{num}}$ denoting the number of heads. The relative positional bias $\mathbf{B} \in \mathbb{R}^{(w_d \cdot w_h \cdot w_w) \times (w_d \cdot w_h \cdot w_w)}$ is a learnable parameter that encodes spatial relationships between tokens within the window, defined as $\mathbf{B}_{ij} = \mathcal{B}(\Delta d, \Delta h, \Delta w)$ where $(\Delta d, \Delta h, \Delta w)$ represents the relative position offset between tokens $i$ and $j$. For shifted windows, an attention mask $\mathcal{M} \in \{0, -\infty\}^{(w_d \cdot w_h \cdot w_w) \times (w_d \cdot w_h \cdot w_w)}$ is added to prevent attention between non-adjacent regions created by cyclic shifting.

Finally, window outputs are merged and the shift is reversed:
\[
\mathbf{Y} = \mathrm{CyclicShift}^{-1}(\mathrm{WindowReverse}(\{\mathbf{Z}^{(m)}\}), \vec{s})
\]
where \(\mathbf{Z}^{(m)}\) denotes the output feature tensor of the \(m\)-th window after W-MSA has been applied, and $\mathrm{WindowReverse}$ concatenates the $M$ window features back into the original spatial layout $\mathbf{Y} \in \mathbb{R}^{B \times D \times H \times W \times C}$. The complete transformer block follows a residual structure:
\begin{align}
\mathbf{X}^{\ell+1/2} &= \mathrm{W\text{-}MSA}(\mathrm{LayerNorm}(\mathbf{X}^{\ell})) + \mathbf{X}^{\ell} \\
\mathbf{X}^{\ell+1} &= \mathrm{MixFFN3D}(\mathrm{LayerNorm}(\mathbf{X}^{\ell+1/2})) + \mathbf{X}^{\ell+1/2}
\end{align}
where $\ell$ denotes the block index, and DropPath regularization is applied to both residual connections during training. This alternating shifted window scheme encourages cross-window information flow while maintaining linear complexity $\mathcal{O}(D \cdot H \cdot W \cdot C)$ with respect to input size, compared to $\mathcal{O}(D^2 \cdot H^2 \cdot W^2 \cdot C)$ for global self-attention. The attention outputs at each stage are then processed through MixFFN3D module to refine and enrich the learned feature representations before downsampling to the next hierarchical level.

\subsubsection{MixFFN3D}

Standard transformer FFNs expand features to $4d$ intermediate dimensions, requiring $8d^2$ parameters per block. We adopt MixFFN~\citep{xie2021segformer} with architectural adaptations for 3D medical volumes. This is done by low-rank factorization reducing channel expansion overhead and 3D depthwise convolution to capture volumetric spatial context within the bottleneck representation.

Given input features $\mathbf{x} \in \mathbb{R}^{B \times N \times d}$, the feed-forward operation factorizes through a low-dimensional subspace:

\begin{equation}
\begin{aligned}
\text{MixFFN3D}(\mathbf{x}) 
&= \mathbf{W}_B \Big( 
\text{SiLU}(\mathbf{W}_A \mathbf{x}) \\
&\quad + \text{DWConv3D}\big(
\text{SiLU}(\mathbf{W}_A \mathbf{x})
\big) 
\Big)
\end{aligned}
\end{equation}

where $\mathbf{W}_A \in \mathbb{R}^{d \times r}$ projects to intermediate rank $r = \max(d/2, 64)$, and $\mathbf{W}_B \in \mathbb{R}^{r \times d}$ projects back to the original dimension. Instead of a single dense layer $\mathbf{W} \in \mathbb{R}^{d \times 4d}$ as in standard FFNs, the low-rank factorization $\mathbf{W}_A \mathbf{W}_B$ approximates the transformation through a bottleneck of rank $r \ll 4d$, where the matrix product $\mathbf{W}_A \mathbf{W}_B \in \mathbb{R}^{d \times d}$ has at most rank $r$.

The 3D depthwise convolution (kernel size $3^3$, applied to reshaped spatial layout) enables local feature aggregation across depth, height, and width within the compressed representation, which is particularly effective for capturing anatomical continuity in medical volumes.

This design reduces parameters from $8d^2$ to $2dr + 27r$. For $d{=}256$ and $r{=}128$, this yields 68,992 parameters versus 524,288 for standard FFN. This gives us a 7.6$\times$ reduction while maintaining expressiveness through the nonlinear spatial mixing operation.

For robust and stable training across diverse data regimes, we employ LayerNorm~\citep{wu2024role} for sequence inputs. Both normalization techniques demonstrate greater stability than BatchNorm~\citep{lubana2021beyond}, especially at small batch sizes commonly found in medical imaging. Additionally, stochastic depth (DropPath)~\citep{sto} regularization is applied throughout the encoder.

Given input features \(\mathbf{X} \in \mathbb{R}^{B \times N \times C}\) in sequence format, LayerNorm computes:
\[
\mathrm{LayerNorm}(\mathbf{X}) = \frac{\mathbf{X} - \mu_\ell}{\sigma_\ell} \cdot \gamma_\ell + \beta_\ell
\]
where \(\mu_\ell\), \(\sigma_\ell\) are the mean and standard deviation computed across all channels \(C\) for each token independently, and \(\gamma_\ell\), \(\beta_\ell\) are learnable scale and shift parameters.

For regularization, we apply DropPath during training:
\[
\mathrm{DropPath}(\mathbf{Z}) =
\begin{cases}
\frac{\mathbf{Z}}{1 - p}, & \text{with probability } 1-p \\
0, & \text{with probability } p
\end{cases}
\]
where \(p\) is the drop probability, typically increased linearly with network depth. This hybrid normalization and regularization scheme enhances training stability and generalization under small batch constraints and limited labeled data.

\subsection{Decoder Architecture}

The decoder progressively reconstructs segmentation maps through three upsampling stages where they incorporate skip connections from corresponding encoder levels. Effective fusion of encoder and decoder features bridges the semantic gap between abstract encoder representations and spatially refined decoder features. Each decoder stage consists of three operations: (i) trilinear upsampling to match encoder resolution, (ii) skip connection fusion via cross attention and (iii) spatial refinement through GhostConv3D blocks.

Each decoder stage follows a progressive refinement pipeline where decoder features 
are first upsampled 2× using trilinear interpolation to match the spatial 
resolution of the corresponding encoder skip connection. The upsampled decoder 
features are then fused with encoder skip connections through window-based 
cross-attention with asymmetric query-key-value assignment. As the decoder features 
generate queries while encoder features provide keys and values, enabling 
selective aggregation of multi-scale encoder context. Finally, the fused 
features undergo spatial refinement through parameter-efficient GhostConv3D 
blocks followed by GroupNorm and SiLU activation, producing enriched 
representations for the subsequent decoder stage or final segmentation head.

Standard U-Net architectures concatenate encoder features $\mathbf{X}_e \in \mathbb{R}^{B \times C_e \times D \times H \times W}$ with decoder features $\mathbf{X}_d \in \mathbb{R}^{B \times C_d \times D \times H \times W}$, treating all encoder information equally regardless of its relevance to the current decoding stage. This uniform fusion strategy assumes that all multi-scale encoder features contribute equally to reconstruction which ignores the semantic context of the decoder's representational state. We instead employ window-based cross-attention for skip connection fusion, where decoder features generate queries to selectively aggregate relevant encoder context, enabling adaptive multi-scale feature integration.

The decoder features $\mathbf{X}_d \in \mathbb{R}^{B \times C_{d} \times D/2 \times H/2 \times W/2}$ from the previous stage are upsampled using trilinear interpolation with scale factor 2:
\begin{equation}
\mathbf{X}_d^{\uparrow} = \mathrm{Upsample}(\mathbf{X}_d, \text{scale}=2) \in \mathbb{R}^{B \times C_d \times D \times H \times W}
\end{equation}
To ensure channel compatibility, encoder skip features are projected via a linear transformation:
\begin{equation}
\mathbf{X}_e' = \mathbf{X}_e \mathbf{W}_{\text{proj}} + \mathbf{b}_{\text{proj}} \in \mathbb{R}^{B \times C_d \times D \times H \times W}
\end{equation}
where $\mathbf{W}_{\text{proj}} \in \mathbb{R}^{C_e \times C_d}$ and $\mathbf{b}_{\text{proj}} \in \mathbb{R}^{C_d}$ are learnable parameters. This projection unifies the channel dimensions to $C_d$ for both streams before attention computation.

After flattening spatial dimensions, $\tilde{\mathbf{X}}_d^{\uparrow}, \tilde{\mathbf{X}}_e' \in \mathbb{R}^{B \times N \times C_d}$ are obtained, where $N = D \times H \times W$ represents the total number of voxels. Decoder features project to queries while encoder features project to key-value pairs:

\begin{align}
\mathbf{Q} &= \tilde{\mathbf{X}}_d^{\uparrow} \mathbf{W}_Q \in \mathbb{R}^{B \times N \times d_k} \\
\mathbf{K}, \mathbf{V} &= \text{Split}(\tilde{\mathbf{X}}_e' \mathbf{W}_{KV}) \in \mathbb{R}^{B \times N \times d_k}
\end{align}
where $\mathbf{W}_Q \in \mathbb{R}^{C_d \times d_k}$ and $\mathbf{W}_{KV} \in \mathbb{R}^{C_d \times 2d_k}$ are learnable projections, with $d_k = C_d / H_{\text{num}}$ denoting the dimension per attention head and $H_{\text{num}}$ the number of heads. The asymmetric query-key-value assignment allows decoder voxels to attend to relevant encoder context through learned similarity, effectively weighting encoder features based on their relevance to each decoder position.

Full attention over $N$ voxels requires $\mathcal{O}(N^2)$ complexity, which is computationally prohibitive for 3D medical volumes (e.g., $N \approx 128^3 \approx 2M$ voxels). We partition features into non-overlapping windows of size $w^3$ (typically $w=4$), yielding $M = \lceil D/w \rceil \times \lceil H/w \rceil \times \lceil W/w \rceil$ windows. Within each window $m \in \{1, \ldots, M\}$, we extract corresponding tokens $\mathbf{Q}_m, \mathbf{K}_m, \mathbf{V}_m \in \mathbb{R}^{B \times w^3 \times d_k}$ and compute multi-head cross-attention:
\begin{align}
\mathbf{A}_m &= \text{Softmax}\left(\frac{\mathbf{Q}_m \mathbf{K}_m^\top}{\sqrt{d_k}}\right) \in \mathbb{R}^{B \times w^3 \times w^3} \\
\mathbf{F}_m &= \mathbf{A}_m \mathbf{V}_m \in \mathbb{R}^{B \times w^3 \times d_k}
\end{align}
The attention matrix $\mathbf{A}_m[i,j]$ encodes the similarity between decoder token $i$ and encoder token $j$ within window $m$, where high values indicate that encoder position $j$ provides relevant context for decoder position $i$. We apply output projection $\mathbf{W}_O \in \mathbb{R}^{d_k \times C_{\text{out}}}$ and reverse window partitioning:

\begin{equation}
\begin{aligned}
\mathbf{Y} &= \text{Reshape}\Big(
\text{WindowReverse}(\{\mathbf{F}_m \mathbf{W}_O\}_{m=1}^M)
\Big) \\
&\in \mathbb{R}^{B \times C_{\text{out}} \times D \times H \times W}
\end{aligned}
\end{equation}

where $\text{WindowReverse}$ concatenates the $M$ window outputs back to the spatial layout. This reduces computational complexity to $\mathcal{O}(N \cdot w^3)$, which is linear in the number of voxels $N$ and enables efficient processing of high-resolution 3D volumes.

Following cross-attention fusion, we apply Squeeze-Excitation (SE) channel attention~\citep{x} to recalibrate channel-wise feature responses:
\begin{align}
\mathbf{z} &= \text{GlobalAvgPool3D}(\mathbf{Y}) \in \mathbb{R}^{B \times C_{\text{out}}} \\
\mathbf{s} &= \sigma(\mathbf{W}_2 \cdot \text{ReLU}(\mathbf{W}_1 \mathbf{z})) \in \mathbb{R}^{B \times C_{\text{out}}} \\
\mathbf{Y}_{\text{se}} &= \mathbf{s} \odot \mathbf{Y}
\end{align}
where $\mathbf{W}_1 \in \mathbb{R}^{(C_{\text{out}}/16) \times C_{\text{out}}}$, $\mathbf{W}_2 \in \mathbb{R}^{C_{\text{out}} \times (C_{\text{out}}/16)}$, $\sigma$ denotes the sigmoid function, and $\odot$ represents channel-wise multiplication. The SE block emphasizes informative channels while suppressing less relevant ones. The refined features are then processed through GhostConv3D~\citep{han2020ghostnet}, GroupNorm, and SiLU activation to produce the final decoder output for stage $i$.

\subsection{Training Objectives}

We employ deep supervision with auxiliary losses at decoder blocks 2 and 3 to stabilize training and guide intermediate feature representations. The overall loss function is defined as:
\begin{equation}
\mathcal{L}_{\text{total}} = \mathcal{L}(\hat{\mathbf{y}}, \mathbf{y}) + \lambda \sum_{i=2}^{3} \mathcal{L}(\hat{\mathbf{y}}_i, \mathbf{y})
\label{eq:total_loss}
\end{equation} 

where ${\mathbf{y}}$ is the ground truth, $\hat{\mathbf{y}}$ is the final prediction, $\hat{\mathbf{y}}_i$ denotes auxiliary predictions from intermediate decoder stages, and $\lambda$  is the weight of the auxiliary supervision. Each loss $\mathcal{L}$ combines Dice loss and cross-entropy loss:
\begin{equation}
\mathcal{L}(\hat{\mathbf{y}}, \mathbf{y}) = \mathcal{L}_{\text{Dice}}(\hat{\mathbf{y}}, \mathbf{y}) + \mathcal{L}_{\text{CE}}(\hat{\mathbf{y}}, \mathbf{y})
\end{equation}

Dice loss addresses class imbalance and directly measures segmentation quality while cross-entropy provides strong gradients for effective optimization. Auxiliary predictions are upsampled to ground truth resolution via trilinear interpolation before computing losses.

\section{Experimental Results}
\label{sec:exp-results}
This section presents the experimental setup and quantitative evaluations conducted to assess the effectiveness and robustness of the proposed method compared to the state-of-the-art methods.
\subsection{Experimental Setup}

We evaluate RefineFormer3D against state-of-the-art 3D medical image segmentation methods on three widely used volumetric benchmarks: BraTS (Brain Tumor Segmentation)~\citep{4b589b6824a64a2a91e8e3b26cc0bf9e} and ACDC (Automatic Cardiac Diagnosis Challenge)~\citep{acdc}. To ensure comparability with prior work, we replicate established protocols for dataset splits, preprocessing, and evaluation. No external data sources, pretraining, or auxiliary datasets are used, and all experiments are performed on a single NVIDIA RTX 5080 GPU using the PyTorch framework.

The model parameters are optimized using the AdamW optimizer~\citep{loshchilov2019decoupledweightdecayregularization}, with an initial learning rate of \(2\times10^{-4}\), weight decay of \(1\times10^{-5}\), and default betas (0.9, 0.999). After warm-up, a cosine annealing schedule~\citep{cos} is employed with a minimum learning rate of \(1\times10^{-9}\) and \(T_{\text{max}}\) set to 100 epochs, unless otherwise specified. The value of $\lambda$(Eq.~\ref{eq:total_loss}) is chosen as 0.4. For ablation studies, ReduceLROnPlateau is optionally used to adapt the learning rate based on validation loss.

To ensure robust generalization, we apply 3D data augmentation including random flipping, rotation, and Gaussian noise. During inference, we utilize test-time augmentation (TTA) via spatial flips and predictions are averaged for final output. All models are evaluated using identical data processing, augmentation, and inference strategies.

\subsection{Quantitative Results}

Table~\ref{table:acdc} presents the performance comparison on the ACDC dataset. RefineFormer3D (GhostConv3D variant, 2.94M parameters) achieves an average Dice score of 93.44\%, outperforming the best competing method, DS-UNETR++ (93.03\%, 67.7M), despite using approximately 95.7\% fewer parameters. The standard Conv3D variant further improves performance to 94.88\% with 4.87M parameters, exceeding DS-UNETR++ by 1.85\%. 

Notably, RefineFormer3D shows consistent gains across all anatomical structures, achieving 92.19\% (RV), 91.97\% (Myo), and 96.14\% (LV) with the lightweight configuration. Compared to transformer-heavy baselines such as nnFormer (150.5M) and TransUNet (96.07M), our model reduces parameter count by over 95–98\% while maintaining superior or competitive segmentation accuracy. This demonstrates exceptional performance and parameter efficiency in cardiac structure segmentation.


Table~\ref{tab:brats_comparison} presents quantitative comparisons on the BraTS dataset. RefineFormer3D achieves an average Dice of 85.9\% (GhostConv3D, 2.94M) and 86.2\% (standard Conv3D, 4.87M). The lightweight variant performs within 0.5\% of nnFormer (86.4\%, 150.5M) while using approximately 98\% fewer parameters, demonstrating a superior accuracy–complexity trade-off. Across tumor subregions, the GhostConv3D variant attains 91.5\% (WT), 80.6\% (ET), and 85.2\% (TC), showing strong whole-tumor and tumor-core segmentation performance under extreme parameter constraints. Even when compared to mid-sized models such as SegFormer3D (4.5M) and GCI-Net (13.36M), RefineFormer3D achieves higher or competitive Dice scores with substantially reduced model capacity.

Overall, the results across both cardiac and brain tumor datasets validate that RefineFormer3D delivers state-of-the-art or near state-of-the-art segmentation accuracy while maintaining an order-of-magnitude reduction in parameters. This establishes the proposed architecture as a highly efficient alternative to conventional transformer-based 3D segmentation frameworks.

\begin{table*}[!t]
\caption{Performance comparison on ACDC dataset. Bold values represent best performance and underlined values indicate second best results. Parameters are reported in millions.}
\label{table:acdc}
\centering
\small
\setlength{\tabcolsep}{8pt}        
\renewcommand{\arraystretch}{1.2}  
\begin{tabular}{@{}lccccc@{}}
\toprule
\textbf{Method} & \textbf{Params (M)} & \textbf{Avg (\%)} & \textbf{RV} & \textbf{Myo} & \textbf{LV} \\
\midrule
DS-UNETR++~\citep{Jiang2025DSUNETRPP} & 67.7 & 93.03 & \underline{92.23} & 90.82 & 96.04 \\
PFormer~\citep{GAO2025107154} & 46.04 & 92.33 & 91.11 & 89.93 & 95.88 \\
nnFormer~\citep{nnFormer} & 150.5 & 92.06 & 90.94 & 89.58 & 95.65 \\
GCI-Net~\citep{10778607} & 13.36 & 91.43 & 90.28 & 89.24 & 94.77 \\
MS-TCNet~\citep{AO2024108057} & 59.49 & 91.43 & 89.43 & 89.09 & 95.77 \\
Segformer3D~\citep{10678245} & 4.5 & 90.96 & 88.50 & 88.86 & 95.53 \\
LeViT-Unet-384~\citep{LeViT-UNet} & 52.17 & 90.32 & 89.55 & 87.64 & 93.76 \\
SwinUNet~\citep{swinUNet} & -- & 90.00 & 88.55 & 85.62 & 95.83 \\
TransUNet~\citep{chen2021transunettransformersmakestrong} & 96.07 & 89.71 & 88.86 & 85.54 & 95.73 \\
DAUNet~\citep{LIU202558} & 16.36 & 88.73 & 85.44 & 86.69 & 94.05 \\
UNETR~\citep{UNETR} & 92.49 & 88.61 & 85.29 & 86.52 & 94.02 \\
R50-VIT-CUP~\citep{chen2021transunettransformersmakestrong} & 86.00 & 87.57 & 86.07 & 81.88 & 94.75 \\
VIT-CUP~\citep{chen2021transunettransformersmakestrong} & 86.00 & 81.45 & 81.46 & 70.71 & 92.18 \\
\midrule
\textbf{RefineFormer3D (GhostConv3D)} & \textbf{2.94} & \underline{93.44} & 92.19 & \underline{91.97} & \underline{96.14} \\
\textbf{RefineFormer3D (Standard Conv3D)} & 4.87 & \textbf{94.88} & \textbf{93.09} & \textbf{93.67} & \textbf{97.89}  \\
\bottomrule
\end{tabular}

\medskip
\footnotesize RV: Right Ventricle; Myo: Myocardium; LV: Left Ventricle.
\end{table*}

\begin{table*}[!t]
\caption{Performance comparison on BraTS dataset. Bold values represent best performance and underlined values indicate second-best results. Parameters are reported in millions.}
\label{tab:brats_comparison}
\centering
\footnotesize
\setlength{\tabcolsep}{16pt}
\begin{tabular}{@{}lccccc@{}}
\toprule
\textbf{Method} & \textbf{Params} & \textbf{Avg} & \textbf{WT} & \textbf{ET} & \textbf{TC} \\
 & \textbf{(M)} & \textbf{(\%)} &  &  &  \\
\midrule
nnFormer~\citep{nnFormer} & 150.5 & \textbf{86.4} & 91.3 & \underline{81.8} & \textbf{86.0} \\
GCI-Net~\citep{10778607} & 13.36 & 85.88 & 91.58 & \textbf{85.86} & 80.20 \\
MS-TCNet~\citep{AO2024108057} & 59.49 & 85.20 & 91.20 & 80.20 & 84.20 \\
DS-UNETR++~\citep{Jiang2025DSUNETRPP} & 67.7 & 83.19 & \underline{91.68} & 78.58 & 79.30 \\
Segformer3D~\citep{10678245} & 4.5 & 82.1 & 89.9 & 74.2 & 82.2 \\
UNETR~\citep{UNETR} & 92.49 & 71.1 & 78.9 & 58.5 & 76.1 \\
TransBTS~\citep{TransBTS} & -- & 69.6 & 77.9 & 57.4 & 73.5 \\
CoTr~\citep{xie2021cotrefficientlybridgingcnn} & 41.9 & 68.3 & 74.6 & 55.7 & 74.8 \\
CoTr w/o CNN~\citep{xie2021cotrefficientlybridgingcnn} & -- & 64.4 & 71.2 & 52.3 & 69.8 \\
TransUNet~\citep{chen2021transunettransformersmakestrong} & 96.07 & 64.4 & 70.6 & 54.2 & 68.4 \\
SETR MLA~\citep{DBLP:journals/corr/abs-2012-15840} & 310.5 & 63.9 & 69.8 & 55.4 & 66.5 \\
SETR PUP~\citep{DBLP:journals/corr/abs-2012-15840} & 318.31 & 63.8 & 69.6 & 54.9 & 67.0 \\
SETR NUP~\citep{DBLP:journals/corr/abs-2012-15840} & 305.67 & 63.7 & 69.7 & 54.4 & 66.9 \\
\midrule
\textbf{RefineFormer3D (GhostConv3D)} & \textbf{2.94} & 85.9 & 91.5 & 80.6 & 85.2 \\
\textbf{RefineFormer3D (Standard Conv3D)} & 4.87 & \underline{86.2} & \textbf{91.8} & {81.3} & \underline{85.6} \\
\bottomrule
\end{tabular}

\medskip
\footnotesize WT: Whole Tumor; ET: Enhancing Tumor; TC: Tumor Core.
\end{table*}

\subsection{Qualitative Analysis}

\begin{figure}[htbp]
\centering

\begin{minipage}{0.30\linewidth}
  \centering
  \small\textbf{Original MRI}
\end{minipage}\hfill
\begin{minipage}{0.30\linewidth}
  \centering
  \small\textbf{Ground Truth}
\end{minipage}\hfill
\begin{minipage}{0.30\linewidth}
  \centering
  \small\textbf{Predicted}
\end{minipage}

\vspace{0.2cm}

\begin{minipage}{0.30\linewidth}
  \centering
  \includegraphics[angle=-90, width=0.75\linewidth]{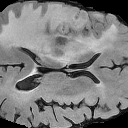}
\end{minipage}\hfill
\begin{minipage}{0.30\linewidth}
  \centering
  \includegraphics[angle=-90, width=0.75\linewidth]{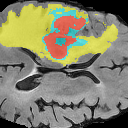}
\end{minipage}\hfill
\begin{minipage}{0.30\linewidth}
  \centering
  \includegraphics[angle=-90, width=0.75\linewidth]{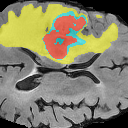}
\end{minipage}

\vspace{0.15cm}

\begin{minipage}{0.30\linewidth}
  \centering
  \includegraphics[angle=-90, width=0.75\linewidth]{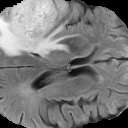}
\end{minipage}\hfill
\begin{minipage}{0.30\linewidth}
  \centering
  \includegraphics[angle=-90, width=0.75\linewidth]{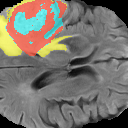}
\end{minipage}\hfill
\begin{minipage}{0.30\linewidth}
  \centering
  \includegraphics[angle=-90, width=0.75\linewidth]{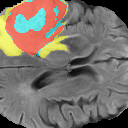}
\end{minipage}

\vspace{0.15cm}

\begin{minipage}{0.30\linewidth}
  \centering
  \includegraphics[angle=-90, width=0.75\linewidth]{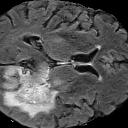}
\end{minipage}\hfill
\begin{minipage}{0.30\linewidth}
  \centering
  \includegraphics[angle=-90, width=0.75\linewidth]{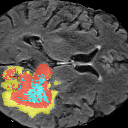}
\end{minipage}\hfill
\begin{minipage}{0.30\linewidth}
  \centering
  \includegraphics[angle=-90, width=0.75\linewidth]{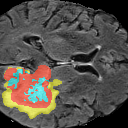}
\end{minipage}

\vspace{0.15cm}

\begin{minipage}{0.30\linewidth}
  \centering
  \includegraphics[angle=-90, width=0.75\linewidth]{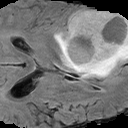}
\end{minipage}\hfill
\begin{minipage}{0.30\linewidth}
  \centering
  \includegraphics[angle=-90, width=0.75\linewidth]{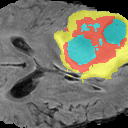}
\end{minipage}\hfill
\begin{minipage}{0.30\linewidth}
  \centering
  \includegraphics[angle=-90, width=0.75\linewidth]{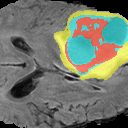}
\end{minipage}

\vspace{0.2cm}

\scriptsize
\centering
\begin{tabular}{@{}c@{\hspace{0.4cm}}c@{\hspace{0.4cm}}c@{}}
\fcolorbox{black}{yellow}{\rule{0.25cm}{0.25cm}} Whole Tumor &
\fcolorbox{black}{orange}{\rule{0.25cm}{0.25cm}} Enhancing Tumor &
\fcolorbox{black}{cyan}{\rule{0.25cm}{0.25cm}} Tumor Core
\end{tabular}

\vspace{0.1cm}

\caption{Visual comparison of original MRI, ground truth, and predicted segmentation for four representative BraTS cases showing whole tumor (yellow), enhancing tumor (orange), and tumor core (cyan) regions.}
\label{fig:brats_results}
\end{figure}

\begin{figure}[htbp]
\centering

\begin{minipage}{0.30\linewidth}
  \centering
  \small\textbf{Original MRI}
\end{minipage}\hfill
\begin{minipage}{0.30\linewidth}
  \centering
  \small\textbf{Ground Truth}
\end{minipage}\hfill
\begin{minipage}{0.30\linewidth}
  \centering
  \small\textbf{Predicted}
\end{minipage}

\vspace{0.2cm}

\begin{minipage}{0.30\linewidth}
  \centering
  \includegraphics[width=0.85\linewidth]{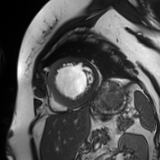}
\end{minipage}\hfill
\begin{minipage}{0.30\linewidth}
  \centering
  \includegraphics[width=0.85\linewidth]{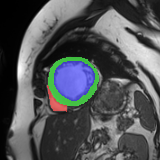}
\end{minipage}\hfill
\begin{minipage}{0.30\linewidth}
  \centering
  \includegraphics[width=0.85\linewidth]{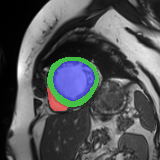}
\end{minipage}

\vspace{0.15cm}

\begin{minipage}{0.30\linewidth}
  \centering
  \includegraphics[width=0.85\linewidth]{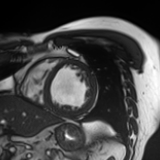}
\end{minipage}\hfill
\begin{minipage}{0.30\linewidth}
  \centering
  \includegraphics[width=0.85\linewidth]{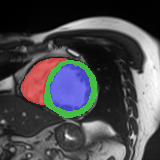}
\end{minipage}\hfill
\begin{minipage}{0.30\linewidth}
  \centering
  \includegraphics[width=0.85\linewidth]{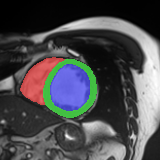}
\end{minipage}

\vspace{0.15cm}

\begin{minipage}{0.30\linewidth}
  \centering
  \includegraphics[width=0.85\linewidth]{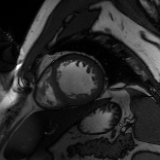}
\end{minipage}\hfill
\begin{minipage}{0.30\linewidth}
  \centering
  \includegraphics[width=0.85\linewidth]{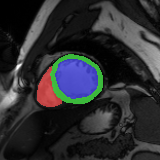}
\end{minipage}\hfill
\begin{minipage}{0.30\linewidth}
  \centering
  \includegraphics[width=0.85\linewidth]{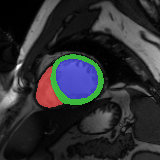}
\end{minipage}

\vspace{0.15cm}

\begin{minipage}{0.30\linewidth}
  \centering
  \includegraphics[width=0.85\linewidth]{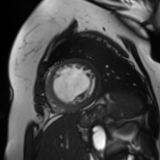}
\end{minipage}\hfill
\begin{minipage}{0.30\linewidth}
  \centering
  \includegraphics[width=0.85\linewidth]{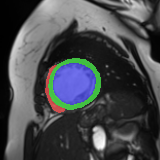}
\end{minipage}\hfill
\begin{minipage}{0.30\linewidth}
  \centering
  \includegraphics[width=0.85\linewidth]{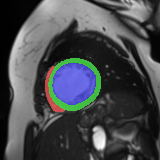}
\end{minipage}

\vspace{0.2cm}

\scriptsize
\centering
\begin{tabular}{@{}c@{\hspace{0.4cm}}c@{\hspace{0.4cm}}c@{}}
\fcolorbox{black}{red!70}{\rule{0.25cm}{0.25cm}} RV &
\fcolorbox{black}{green!70}{\rule{0.25cm}{0.25cm}} Myocardium &
\fcolorbox{black}{blue!70}{\rule{0.25cm}{0.25cm}} LV
\end{tabular}

\vspace{0.1cm}

\caption{Visual comparison of original MRI, ground truth, and predicted segmentation for four representative ACDC cardiac cases showing right ventricle (RV), myocardium, and left ventricle (LV).}
\label{fig:acdc_results}
\end{figure}

To qualitatively assess segmentation performance, Figures~\ref{fig:brats_results} and~\ref{fig:acdc_results} present visual comparisons of predicted segmentations with ground truth annotations on the BraTS and ACDC datasets, respectively.

Figure~\ref{fig:brats_results} presents segmentation outcomes on four representative cases from the BraTS dataset. Each row shows an original axial slice from multimodal brain MRI scans, followed by ground truth annotation and predicted segmentation. The results demonstrate effective identification and distinction of tumor subregions, including small enhancing areas and complex structures. Predicted boundaries align well with ground truth, even in regions with irregular shapes and heterogeneous intensities.

Figure~\ref{fig:acdc_results} illustrates segmentation results on four representative ACDC cases. Each row shows an original cardiac MRI slice, followed by corresponding ground truth segmentation and predicted segmentation. The predicted mask shows close similarity with ground truth majorly around LV and myocardium edges. The model effectively captures detailed boundaries and adapts well to different anatomical variations. This demonstrates strong potential for clinical applications where reliable segmentation is essential.

\subsection{Robustness to Reduced Training Data}

To test generalization under limited supervision, we progressively reduce the BraTS training set to 90\%, 70\%, and 50\% of the original cases while keeping the test set unchanged. As presented in Table~\ref{tab:ablation-data}, RefineFormer3D maintains stable performance across all settings: 85.90\% (100\%), 85.80\% (90\%), 84.86\% (70\%), and 82.50\% (50\%). The performance degradation remains modest even under severe data reduction. Reducing the training data to 70\% results in only a 0.94 point  decrease in Dice score, while halving the dataset (50\%) leads to a 3.40 point absolute drop. This gradual decline indicates strong generalization capability and resistance to overfitting.

The observed robustness achieved on the reduced training dataset is due to the compact architectural design, which limits over-parameterization while preserving effective contextual modeling. In the medical imaging domain, where annotated volumetric data are scarce and costly to obtain, such stability under reduced supervision is particularly valuable. These findings show that RefineFormer3D provides favorable regularization characteristics.

\begin{table}[!hbt]
\caption{Robustness of RefineFormer3D to reduced training data on BraTS.}
\label{tab:ablation-data}
\centering
\begin{tabular}{@{}lcc@{}}
\toprule
\textbf{\% Training Set} & \textbf{Dice (\%)} & \textbf{Drop in Dice} \\
\midrule
100\% & \textbf{85.90} & -- \\
90\%  & 85.80 & 0.10 \\
70\%  & 84.86 & 0.94 \\
50\%  & 82.50 & 3.40 \\
\bottomrule
\end{tabular}
\end{table}

\subsection{Component wise Ablation Analysis}

We conducted ablations of key components on the BraTS dataset to quantify the contribution of each major component- MixFFN3D, Cross Attention Fuion and GhostConv3D. As shown in Table~\ref{tab:ablations}, removing or replacing any of these modules consistently decreases Dice performance, confirming their complementary roles in efficiency and accuracy.

\begin{table*}[!hbt]
\caption{Ablation results on BraTS. Each variant is trained under identical settings. Dice (\%) \(\uparrow\); lower \(\Delta\) indicates performance drop relative to the full model.}
\label{tab:ablations}
\centering
\begin{tabular}{@{}lcc@{}}
\toprule
\textbf{Variant} & \textbf{Dice (\%)} & \textbf{Params (M)} \\
\midrule
\textbf{Full RefineFormer3D} & \textbf{85.90} & \textbf{2.94} \\
w/o MixFFN3D \(\rightarrow\) Dense MLP & 84.87 & 2.88 \\
w/o Cross-Attention Fusion \(\rightarrow\) Concat + Conv3D & 83.22 & 3.00 \\
GhostConv3D \(\rightarrow\) Standard Conv3D & 86.21 & 4.87 \\
\bottomrule
\end{tabular}
\end{table*}

Disabling MixFFN3D or replacing it with a dense MLP leads to a 1.03\% drop in Dice score (85.90\% → 84.87\%) while maintaining similar parameter count (2.94M → 2.88M), validating its low-rank efficiency. Removing the cross attention fusion and using simpler concatenation with Conv3D results in the largest performance degradation of 2.68\% (85.90\% → 83.22\%), demonstrating its critical role in feature fusion. Replacing GhostConv3D with standard Conv3D increases parameters by 66\% (2.94M → 4.87M) while  improving Dice slightly to 86.21\%, indicating that while standard convolutions offer marginally better accuracy, the parameter cost is prohibitive. Together, these ablations confirm that each component is essential for achieving the optimal accuracy-efficiency trade off in RefineFormer3D.

\subsection{Inference Efficiency and Memory Usage}
\label{sec:efficiency}

Table~\ref{tab:brats_efficiency} summarizes inference and memory performance of RefineFormer3D versus transformer-based 3D segmentation baselines. It achieves the best trade-off between speed, accuracy and model size requiring only 2.94\,M parameters and 191.2\,GFLOPs. Despite being nearly \(50\times\) smaller than nnFormer, \(20\times\) smaller than SwinUNETR and \(15\times\) smaller than PFormer, it sustains rapid inference with a forward-pass latency of only 8.35\,ms on GPU and 296.2\,ms on CPU.

\begin{table}[!hbt]
\caption{Inference efficiency on BRaTS input (\(1\times 4\times 128^3\)). Times averaged over 200 passes. Env: RTX 5080, Ryzen 9 9950X, PyTorch 2.8, CUDA 12.8.}
\label{tab:brats_efficiency}
\centering
\footnotesize
\setlength{\tabcolsep}{3pt}
\begin{tabular}{@{}lccccc@{}}
\toprule
\textbf{Model} & \textbf{Params} & \textbf{FLOPs} & \textbf{Mem\(^{\dagger}\)} & \textbf{GPU} & \textbf{CPU} \\
 & \textbf{(M)} & \textbf{(G)} & \textbf{(GB)} & \textbf{(ms)} & \textbf{(ms)} \\
\midrule
UNETR~\citep{UNETR} & 92.5 & 153.5 & 3.3 & 82.5 & 2145 \\
SwinUNETR~\citep{10.1007/978-3-031-08999-2_22} & 62.8 & 572.4 & 19.7 & 228.6 & 7612 \\
nnFormer~\citep{nnFormer} & 149.6 & 421.5 & 12.6 & 148.0 & 5248 \\
DS-UNETR++~\citep{Jiang2025DSUNETRPP} & \underline{42.6} & \textbf{70.1} & \underline{2.4} & \underline{62.4} & \underline{1498} \\
\midrule
\textbf{RefineFormer3D} & \textbf{2.94} & \underline{191.2} & \textbf{1.5} & \textbf{8.35} & \textbf{296} \\
\bottomrule
\end{tabular}

\smallskip
\noindent\scriptsize\textit{\(^{\dagger}\)Peak GPU memory via PyTorch allocator; system peak 2.1\,GB.}
\end{table}

The peak GPU memory footprint is significantly lower than prior transformer baselines, making RefineFormer3D suitable for both workstation and embedded deployment. When normalized by computational cost, RefineFormer3D attains: ms/GFLOP = 0.043,  vol/s/GFLOP = 15.3, \\ mem(GB)/Mparam = 0.51, demonstrating superior compute efficiency and throughput-per-FLOP compared to UNETR++ (0.89, 0.23, 0.056 respectively).

\section{Conclusion}
\label{sec:con}
In this work, we presented RefineFormer3D, a parameter-efficient hierarchical transformer based architecture for 3D medical image segmentation that systematically addresses the accuracy–efficiency trade-off. By integrating GhostConv3D-based patch embedding, MixFFN3D with low-rank spatial–channel mixing and an adaptive cross-attention skip fusion decoder, the proposed framework achieves strong contextual modeling with minimal computational overhead.
The experiments on the BraTS and ACDC benchmarks demonstrate that RefineFormer3D achieves state-of-the-art or competitive Dice performance while operating with substantially fewer parameters and a reduced memory footprint compared to competitive transformer baselines. By improving throughput and deployment feasibility, RefineFormer3D contributes toward the translation of transformer-based segmentation systems into real-world clinical workflows. Future work will evaluate the model’s ability to generalize across different imaging modalities and adapt to data variability in multi-institutional settings. Additionally, we will investigate its integration into end-to-end computer-assisted diagnosis and clinical decision-support systems.

\section*{Acknowledgments}
The authors would like to thank the researchers who made the BraTS and ACDC datasets publicly available. We also acknowledge the computational resources provided by the Indian Institute of Technology Ropar.

\section*{Declaration of Competing Interest}
The authors declare that they have no known competing financial interests or personal relationships that could have appeared to influence the work reported in this paper.

\section*{Data Availability}
The datasets analyzed during the current study are publicly available:
\begin{itemize}
\item BraTS: \url{https://www.med.upenn.edu/cbica/brats/}
\item ACDC: \url{https://www.creatis.insa-lyon.fr/Challenge/acdc/}
\end{itemize}

\bibliographystyle{plain}
\bibliography{mybibfile}

\end{document}